\documentstyle [prl,aps,epsf]{revtex}
\begin{document}
\twocolumn[{
\widetext
\draft

\title{
Incoherent interlayer transport and angular-dependent\\
 magnetoresistance oscillations in layered metals
}

\author{Ross H. McKenzie\cite{email}
 and Perez Moses}

\address{School of Physics, University of New
South Wales, Sydney 2052, Australia}

\date{Received 10 June 1998}
\maketitle
\mediumtext
\begin{abstract}
The effect of incoherent interlayer transport
on the interlayer resistance of a layered metal is considered.
We find that for both quasi-one-dimensional
and quasi-two-dimensional Fermi liquids
the angular dependence of the magnetoresistance
is essentially the same for coherent and incoherent transport.
Consequently, the existence of a three-dimensional
Fermi surface is {\it not} necessary to explain
the oscillations in the magnetoresistance
that are seen in many organic conductors
as the field direction is varied.
\\
\\
Published as Phys. Rev. Lett. {\bf 81}, 4492 (1998).
\\
\end{abstract}

%\pacs{PACS numbers: 74.70.Kn, 72.15.Gd, 71.10.Hf, 74.20.Mn}

}]
\narrowtext

One of the most fundamental concepts in
solid state physics is that in most metallic crystals
the electronic conduction occurs through the
coherent motion of electrons in band states
associated with well-defined wave vectors\cite{ashcroft}.
There is currently a great deal of interest
in whether this concept is valid for interlayer
transport in high-$T_c$ superconductors\cite{anderson,hussey2},
organic conductors\cite{str}, and layered manganite
compounds with colossal magnetoresistance\cite{kimura}.
Incoherent transport means that the motion from
layer to layer is diffusive and band states 
and a Fermi velocity perpendicular to the
layers cannot be defined.
The Fermi surface is then not three-dimensional and Boltzmann
transport theory cannot describe the interlayer transport.

In organic conductors\cite{ish} large variations
in the magnetoresistance are observed as the direction
of the magnetic field is varied and are referred to
as angular-dependent magnetoresistance oscillations
(AMRO)\cite{wos}.
These effects
in quasi-one-dimensional systems
are known as Danner\cite{dan}, Lebed\cite{dan2,dan3,lebed0},
 and third angular effects\cite{third},
depending on whether the magnetic field is rotated
 in the ${\bf a}-{\bf c}$, ${\bf b}-{\bf c}$, or ${\bf a}-{\bf b}$
 plane, respectively. (The ${\bf a}$ and ${\bf c}$ axes are
the most- and least-conducting directions, respectively).
Oscillations in quasi-two-dimensional systems
include the Yamaji\cite{yam} oscillations
and the anomalous AMRO in 
 the low-temperature phase of
$\alpha$-(BEDT-TTF)$_2$MHg(SCN)$_4$[M=K,Rb,Tl]\cite{wos,khg}.

We focus on the Danner and Yamaji  oscillations here
because their explaination in terms
of a three-dimensional Fermi surface has generally
been accepted.
 The resistance perpendicular to the layers is a maximum when the
field direction is such that the electron velocity
(perpendicular to the layers) averaged
over its trajectories on the Fermi surface is zero\cite{dan,kart}.
In contrast, it is not clear that
coherent transport models can explain the angle-dependent
magnetoresistance
in  the quasi-one-dimensional (TMTSF)$_2$PF$_6$ at pressures of about 10
kbar\cite{str,dan2,dan3,yakov,naughton}.
The main result of this Letter 
is that coherent interlayer transport is not {\it necessary}
to explain the Yamaji and Danner oscillations.
In contrast, the observation of beats in the magneto-oscillations
of quasi-two-dimensional systems and a peak in the magnetoresistance
when the field is parallel to the layers is 
evidence for a three-dimensional Fermi surface.
We now define precisely
what we mean by coherent and incoherent transport (see Fig. 1)
and how to calculate the associated conductivity.

{\it Coherent interlayer transport.}
A three-dimensional dispersion relation
$\epsilon_{3D}(\vec k)$
 can be defined where 
\begin{equation}
\epsilon_{3D}(\vec k)= \epsilon(k_x,k_y) - 2 t_c \cos (k_z c)
\label{eq:disp}
\end{equation}
where $t_c$ is the interlayer hopping integral,
$c$ is the layer separation, and 
$\epsilon(k_x,k_y)$ is the intra-layer dispersion
relation, simple examples of which are given in Table I.
The electronic group velocity perpendicular to
the layers is
\begin{equation}
v_z = {1 \over \hbar} {\partial \epsilon_{3D} (\vec k) \over \partial k_z}
= { 2 t_c c \over \hbar} \sin(k_z c).
\;
\label{eq:groupvel}
\end{equation}
The interlayer conductivity involves correlations of
this velocity and is given by
Chambers formula\cite{ashcroft}
\begin{equation}
\sigma_{zz} = {e^2 \tau\over 4 \pi^3}
\int d^3 k \ {v_z(\vec k) \bar{v}_z(\vec k)}
\delta(E_F - \epsilon_{3D} (\vec k))
\label{eq:chambers}
\end{equation}
where $E_F$ is the Fermi energy, $\tau$ the scattering time, and
$\bar{v}_z(\vec k)$ is the velocity averaged over a trajectory
on the Fermi surface ending at $\vec k$:
\begin{equation}
\bar{v}_z(\vec k) =
{1 \over \tau}
 \int_{-\infty}^{0} dt
\exp(t / \tau) v_z(\vec k(t)). \; \label{eq:vbar}
\end{equation}
If the magnetic field is tilted
 sufficiently far away from the layers that
$t_c c \tan \theta \ll \hbar v_F$,
where $\theta$ is the angle between the field and the
normal to the layers,
 then to lowest order in $t_c$ the expression
(\ref{eq:chambers}) can be evaluated analytically.
This means neglecting the effects of closed orbits
that become important when the field direction
is close to the layers\cite{hanasaki}.
After long calculations the results 
for both the quasi-one- and quasi-two-dimensional
cases can be written in the form (\ref{amro})
given below.

{\it Incoherent interlayer transport.}
If the intralayer scattering rate $1/\tau$ is much larger
than the interlayer hopping integral $t_c$ \cite{estimate}
then the interlayer transport will be incoherent\cite{nfl}
in the sense that successive interlayer tunnelling
events are uncorrelated\cite{kumar0}.
The interlayer conductivity
is then proportional to the tunnelling rate
between just two adjacent layers (see Fig. 1).
This rate can be  calculated using standard formalisms
for tunneling in metal-insulator-metal junctions\cite{mansky,mahan}
which assume that the intralayer momentum is conserved. 
The result (for temperatures much less than the Fermi
energy and $\hbar = 1$) is
\begin{equation}
\sigma_{zz}  
 =  { e^2 t_{c}^{2} c \over \pi L^2}
 \int{d^2r_a d^2r_b}
 A_{1}(\vec r_a,\vec r_b,E_F) A_{2}(\vec
r_b,\vec r_a,E_F)
\label{sigzz}
\end{equation}
where $L^2$ is the area of the layer
and $A_{j}(\vec r_a,\vec r_b,E) (j=1,2)$ are the spectral functions
 for layers 1 and 2.
It will be seen below that in the presence of a tilted magnetic
field $A_1$ and $A_2$ are not identical.
The zero-field limit of this expression 
has been used in treatments of incoherent interlayer
transport in the cuprate superconductors\cite{kumar}.

The magnetic field $\vec B = (B_x,0,B_z)=
(B\sin \theta ,0, B \cos \theta)$
is described by a vector potential $\vec A$, which in the Landau
gauge has only one non-zero component,
$A_y = B_z x - B_x z$.
The Hamiltonian for layer 1 $(z = 0)$ is then
the same as that for a single layer in
a perpendicular field $B \cos \theta$.
The Hamiltonian for layer 2 $(z = c)$ is the
the same as for layer 1 except
 $x$ is replaced with $(x- c\tan \theta)$.
This displacement actually corresponds to a gauge
transformation\cite{kleinert},
$\vec A \to \vec A - \nabla \Lambda$ where $\Lambda(\vec r)= B \sin \theta c y$.
Wave functions transform according to
$\psi(\vec r) \to \psi(\vec r) \exp ( i e \Lambda(\vec r))$.
The Green's functions in layers 1 and 2 are then related by
\begin{equation}
G_{2}^{}(\vec r_a , \vec r_b) = \exp ( i e \Lambda(\vec r_a))
 G_{1}^{}(\vec r_a , \vec r_b)
\exp (-i e \Lambda(\vec r_b))
\label{g15}
\end{equation}
Substituting this in (\ref{sigzz}) gives 
\begin{equation}
\sigma_{zz}  
 =  {2 e^2 t_{c}^{2} c \over \pi}  \int{d^2r}
 \hspace{4pt} |G_1 (\vec r,0,E_F)|^2 
\cos \left( e B \sin\theta  c \ y \right).
\label{sigzz2}
\end{equation}

We have evaluated (\ref{sigzz2}) for the simplest
possible situation, a Fermi liquid within each
layer, with the dispersion relations
given in Table I.
The complete details of the calculations will be given
elsewhere\cite{moses}.
For the quasi-two-dimensional case we
followed a procedure similar to that used
by Hackenbroich and von Oppen\cite{hack}
in their study of magneto-oscillations in
anti-dot lattices.
In the semi-classical approximation the Green's function
is written as a sum over classical trajectories from
$\vec r_a$ to $\vec r_b$.
For the quasi-one-dimensional case 
the quasi-classical Green's function\cite{gorkov}
was used.

In a tilted magnetic field
the interlayer conductivity for {\it both}
coherent and incoherent interlayer transport is
\begin{equation}
\sigma_{z z}(\theta)
 =\sigma_{z z}^{0} \hspace{1mm} [ J_{0}(\gamma \tan\theta)^2
+ 2
\sum_{\nu=1}^{\infty} { J_{\nu}(\gamma \tan\theta)^2
\over {1 + (\nu \omega_0 \tau \cos \theta)^2} }]
\label{amro}
\end{equation}
where 
$\sigma_{zz}^0$ is the zero-field conductivity,
 $J_{\nu}(x)$ is the $\nu$-th order
Bessel function, $\omega_0$ is the oscillation frequency
associated with the magnetic field, and 
$\gamma $ is a constant that depends on the geometry of
the Fermi surface (see Table I).
This expression was previously derived by Yagi et al.\cite{yagi}
for coherent interlayer transport for a quasi-two-dimensional
Fermi surface\cite{yosh}.
If $\omega_0 \tau \cos \theta \gg 1$ then the first term in (\ref{amro})
is dominant.
However, if $\gamma \tan \theta$ equals a zero of
the zero-th order Bessel function then
at that angle $\sigma_{zz}$ will be a minimum
and the interlayer resistivity will be a maximum.
If $\gamma \tan \theta \gg 1$, then the zeroes occur
at angles $\theta_n$ given by 
\begin{equation}
\gamma \tan \theta_n = \pi (n - {1 \over 4}) \ \ \ \ \ (n=1,2,3, \cdots).
\label{minangle}
\end{equation}
Determination of these angles experimentally provides
a value for $\gamma$ and thus information about the intralayer Fermi surface.
The values of the Fermi surface area of quasi-two-dimensional systems
determined  from AMRO are in good agreement with the
Fermi surface areas determined from the frequency
of magneto-oscillations\cite{wos}.
 
Fig. 2 shows the angular dependence of the interlayer resistivity
$\rho_{zz} \equiv 1/ \sigma_{zz}$
for parameter values relevant to 
(TMTSF)$_2$ClO$_4$.
The results are similar to the experimental results
in Ref. \onlinecite{dan} and the results
of numerical integration of Chambers formula
for coherent transport (\ref{eq:chambers})
except near 90 degrees. For coherent transport there is
a small peak in $\rho_{zz}(\theta)$ at $\theta = 90$ degrees.
This is due to the existence of closed orbits on the Fermi surface when
the field lies close to the plane of the layers\cite{hanasaki}.
For incoherent transport these orbits do not exist
and so the associated magnetoresistance is not present.
Hence, except close to 90 degrees,
the Danner oscillations can be explained equally well in
terms of incoherent transport.
Hence, contrary to the claims of Ref. \onlinecite{dan2},
the observation of
Danner oscillations is not necessarily evidence for the existence of 
a three-dimensional Fermi surface.
Similarly, the suppression of the Danner oscillations by
the introduction of 
a small component of the magnetic field in the ${\bf b}$ direction,
as is observed in 
(TMTSF)$_2$PF$_6$ at pressures of about 10 kbar\cite{dan2},
does not necessarily imply
that the field is destroying the three-dimensional Fermi surface.

It is the averaging of the phase factor over the spatial integral
in (\ref{sigzz2}) that gives rise
to the Yamaji and Danner effects.
The length scale associated with the magnetic field
 for the quasi-2d system is the
cyclotron length $R$ which at the Fermi energy is
$R = \hbar k_F / (e B \cos \theta)$.
For the quasi-1d case the length scale associated
with oscillations perpendicular to the chains
is $ R = 2 t_b/(e v_F B \cos \theta)$\cite{chaikin}.
At this length scale the phase difference between the wave function 
of adjacent layers is $e \Lambda(R) = e B \sin\theta  c R = \gamma \tan
\theta $.
Naively, we might expect maximum resistivity when this phase difference
is an odd multiple of $\pi$, leading to a condition 
different from (\ref{minangle}).
However, one must take into account averaging
of the electron position over the perpendicular direction.

Given we have shown that the existence of a
three-dimensional Fermi surface is not necessary
to produce the Yamaji oscillations we consider an alternative test
for coherent transport for quasi-two-dimensional systems.
Definitive evidence for the existence of a
three-dimensional Fermi surface, such as that shown
in Fig. 1 (a), is the observation of
a beat frequency in de Haas-van Alphen 
and Shubnikov - de Haas oscillations.
The frequency of these oscillations is determined
by extremal areas of the Fermi surface\cite{wos}.
For the Fermi surface shown in Fig. 1 (a)
there are two extremal areas, corresponding to
``neck'' and ``belly'' orbits. The small difference between
the two areas leads to a beating of the corresponding
frequencies with a frequency proportional to $t_c/E_F$\cite{wos}.
Such beat frequencies have 
been observed in $\beta$-(BEDT-TTF)$_2$I$_3$,
$\beta$-(BEDT-TTF)$_2$IBr$_2$\cite{wos},
$\alpha$-(BETS)$_2$KHg(SCN)$_4$ at pressures above 4 kbar\cite{agosta},
 and Sr$_2$RuO$_4$\cite{mack}.
In the former it was used to establish that 
$t_c/E_F \simeq 1/175$\cite{wos}.
However, in many other quasi-two-dimensional
organics no beat frequency is observed\cite{wos}.
This could be because 
 the interlayer transport is incoherent
or  because the interlayer hopping $t_c$ is 
so small that the beat frequency cannot be resolved 
experimentally. For $\kappa$-(BEDT-TTF)$_2$I$_3$
the absence of beating has been used
to establish the upper bound  $t_c/E_F < 1/3000$\cite{wos,balthes}.
This implies a conductivity anisotropy
$\sigma_{zz}/\sigma_{xx} \sim (t_c/E_F)^2 < 10^{-7}$.
However, the observed anisotropy in the
$\kappa$-(BEDT-TTF)$_2$X materials is about
$10^{-3}$\cite{dressel}.
This large discrepancy suggests that the interlayer transport
is incoherent in these materials.

We have also examined semi-classical transport
models\cite{lebed0}
 which give  Lebed resonances and  find that 
the resonances are still present 
for incoherent interlayer transport\cite{moses}.
A much greater challenge than that considered
here is to explain the angle-dependent magnetoresistance observed
in (TMTSF)$_2$PF$_6$ at pressures of about 10 kbar\cite{dan2,dan3}.
In particular, the background magnetoresistance is smallest
when the field is in the layers,
the opposite of what one expects based on
the simple Lorentz force arguments relevant
to semi-classical magnetoresistance.

We thank C. C. Agosta,
P. W. Anderson, J. S. Brooks, P. M. Chaikin, D. G. Clarke,
G. M. Danner, S. Hill, A. H. MacDonald, and
S. P. Strong for helpful discussions.

\vskip -0.5cm

\begin{figure}
\centerline{\epsfxsize=9.3cm \epsfbox{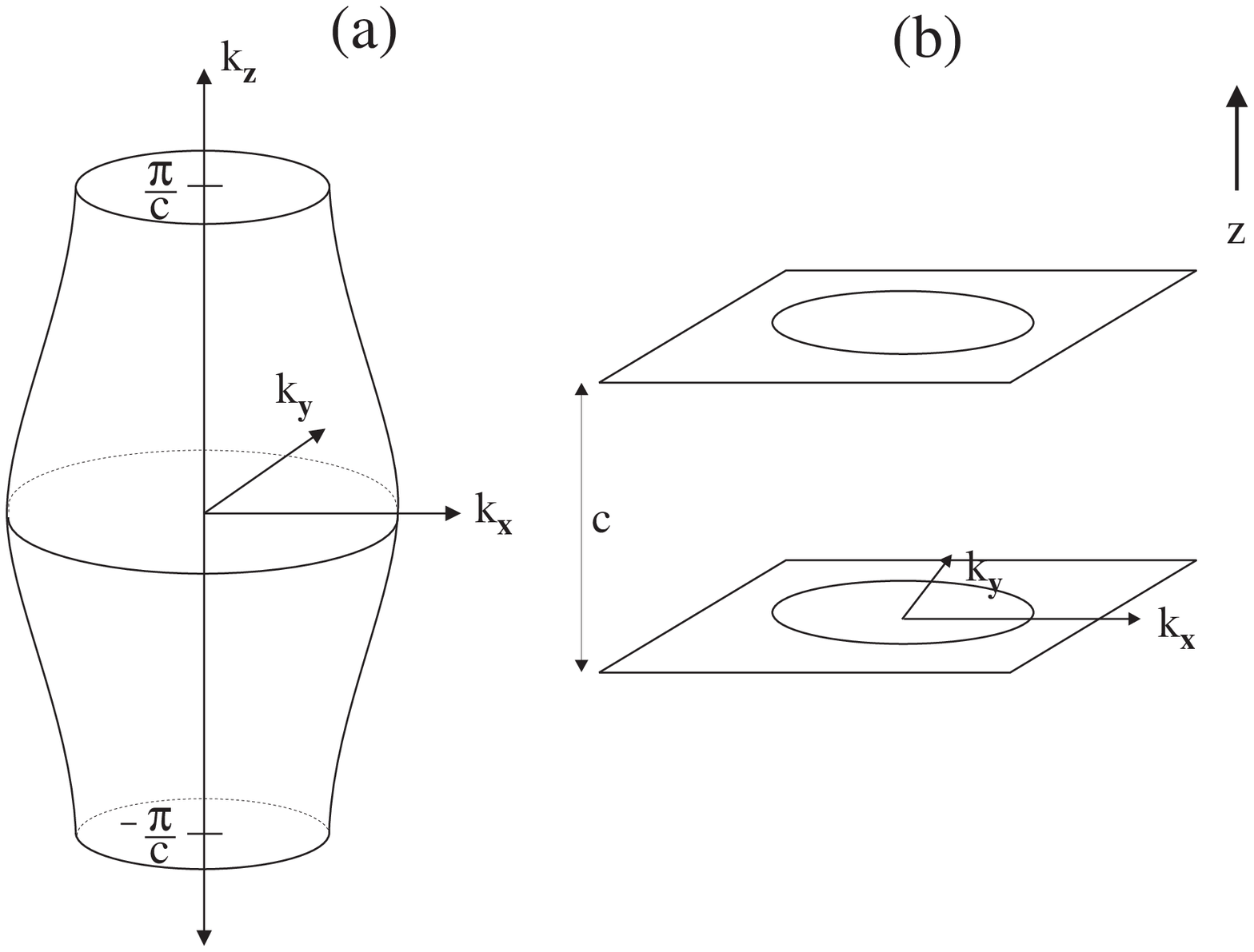}}
\caption{
The pictures relevant to coherent and
incoherent interlayer transport in a quasi-two-dimensional
system.
(a) If the transport between layers is coherent
then one can define a three dimensional Fermi
surface which is a warped cylinder.
The interlayer conductivity is determined by
correlations of the electronic         group velocity 
perpendicular to the layers.
(See equation (\protect\ref{eq:chambers})).
(b) For the incoherent interlayer transport
considered here
a Fermi surface is only
defined within the layers and the interlayer
conductivity is determined by the
interlayer tunnelling rate.
(See equation (\protect\ref{sigzz})).
\label{fig1}}
\end{figure}

\begin{figure}
\centerline{\epsfxsize=9.3cm \epsfbox{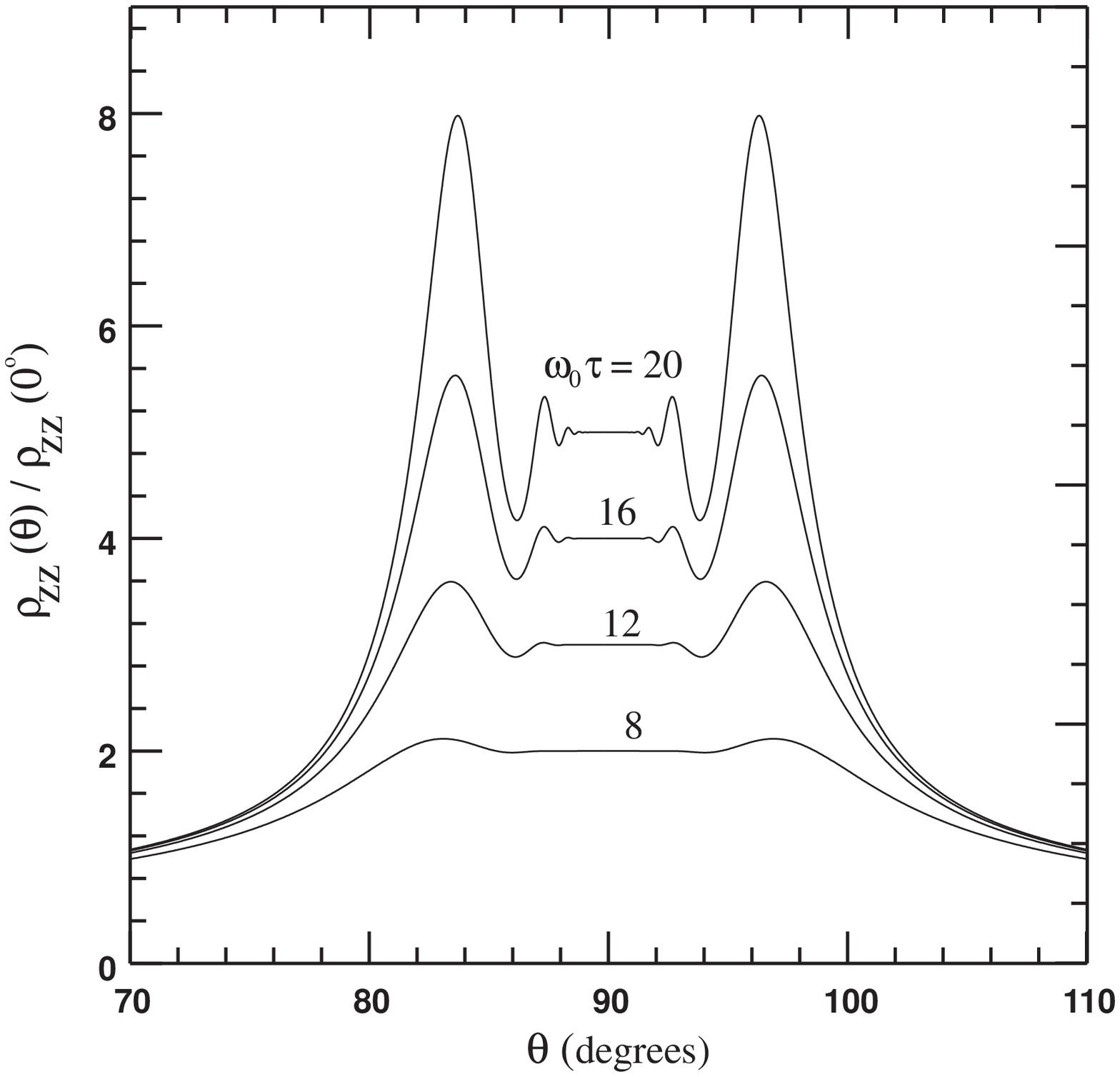}}
\vskip 0.5cm
\caption{
Dependence of the interlayer resistance of
a quasi-one-dimensional system on
the direction of the magnetic field for a range
of magnetic fields.
$\theta$ is the angle between the magnetic field and
the least conducting direction, with the field in the 
same plane as the most conducting direction.
The parameter which defines the anisotropy of
the intralayer hopping $\gamma = 0.25$ (cf. Table I).
$\tau$ is the intralayer scattering time
and $\omega_0$ is the frequency at which the electrons
oscillate between the chains when the field is
perpendicular to the layers.
Except very close to 90 degrees
this figure is similar to the experimental 
data on (TMTSF)$_2$ClO$_4$ in Ref. \protect\onlinecite{dan}.
\label{fig2}} \end{figure}

\newpage
\onecolumn
\widetext

\begin{table}
\caption{
Different physical quantities relevant 
to angular-dependent magnetoresistance oscillations
for the cases where intralayer Fermi surface is
quasi-one-dimensional (open)
and quasi-two-dimensional (closed).
In a magnetic field the electrons oscillate
on the Fermi surface with frequency $\omega_0$
when the field $B$ is
perpendicular to the layers.
The geometric factor $\gamma$ determines the
field directions at which the interlayer resistivity
is a maximum (see equation (\protect\ref{minangle})).
The magnitude of the Fermi wavevector is denoted $k_F$.
For the quasi-one-dimensional case, $v_F$ is the
Fermi velocity, $t_b$ the interchain hopping integral,
and $b$ the interchain distance.
For the quasi-two-dimensional case, $m^*$ is the
effective mass.
}
\begin{tabular}{lccc}
Quantity & Symbol & Quasi-1d  & Quasi-2d \\
\tableline
\vtop{\baselineskip=10 pt \halign{#\hfil \cr
Intra-layer \cr dispersion \cr}}
 & $\epsilon(k_x,k_y)$ &
 $ \displaystyle \hbar v_F (|k_x| - k_F) - 2 t_b \cos (k_y b) $
  & $  \displaystyle {\hbar^2 \over 2 m^*} (k_x^2 + k_y^2) $ \\
\vtop{\baselineskip=10 pt \halign{#\hfil \cr
Oscillation \cr frequency \cr}}
 & $\omega_0$ & $ \displaystyle {e v_F b B   \over \hbar } $ & 
$ \displaystyle {e  B   \over m^*   } $ \\
 \vtop{\baselineskip=10 pt \halign{#\hfil \cr
Geometric  \cr factor \cr}}
 & $\gamma$ & $ \displaystyle{2 t_b c  \over \hbar v_F } $
 & $ k_F c$  \\
 \vtop{\baselineskip=10 pt \halign{#\hfil \cr
Zero-field interlayer \cr conductivity \cr}}
 & $ \sigma_{z z}^{0} $ & 
$\displaystyle { 4e^2  c t_c^2 \tau \over \pi \hbar^3 b v_F }$ & 
$\displaystyle { 2e^2 m^\star c t_c^2 \tau \over \pi \hbar^4}$   \\
\end{tabular} \label{table2}
\end{table}

\begin{references}
\bibitem{estimate}
Previous estimates of $t_c$ and $\tau$
in various organics 
(e.g., \cite{wos,dan,balthes}) suggest
these quantities may be comparable.
\bibitem{nfl}
Alternatively, incoherence may arise
due to non-Fermi liquid effects\cite{anderson,str}.
\end{references}
\end{document}